\documentclass[conference]{IEEEtran}
\IEEEoverridecommandlockouts
\usepackage{amsmath}
\usepackage{float}
\usepackage{enumerate}

\usepackage{pgfplots}
\usepackage{mathtools}
\usepackage{enumerate}
\usepackage{lipsum}

\usepackage{float}
\usepackage{lettrine}
\usepackage{graphicx}
\usepackage{subfigure}
\usepackage{tikz}
\usetikzlibrary{patterns}
\usetikzlibrary{backgrounds,automata}
\usepackage{latexsym}
\usepackage{graphicx}
\usetikzlibrary{decorations.markings}
\usepackage{fancybox}
\begin{document}
\title{Jamming in multiple independent Gaussian\\ channels as a game}

\author{\IEEEauthorblockN{Michail Fasoulakis\IEEEauthorrefmark{1},
Apostolos Traganitis\IEEEauthorrefmark{1}\IEEEauthorrefmark{2},
Anthony Ephremides\IEEEauthorrefmark{3}}

\IEEEauthorblockA{\IEEEauthorrefmark{1}Institute of Computer Science,\\ Foundation for Research and Technology-Hellas (ICS-FORTH), Greece\\
Email: \{mfasoul, tragani\}@ics.forth.gr}
\IEEEauthorblockA{\IEEEauthorrefmark{2}Department of Computer Science, University of Crete, Greece}
\IEEEauthorblockA{\IEEEauthorrefmark{3}Department of Electrical and Computer Engineering, and
Institute for Systems Research,\\
University of Maryland, College Park, USA\\
Email: etony@umd.edu}}\maketitle

\begin{abstract}
We study the problem of \emph{jamming} in multiple independent \emph{Gaussian channels} as a zero-sum game. We show that in the unique Nash equilibrium of the game the best-response strategy of the transmitter is the \emph{waterfilling} to the sum of the jamming and the noise power in each channel and the best-response strategy of the jammer is the \emph{waterfilling} only to the noise power.
\end{abstract}

\section{Introduction}
One of the fundamental problems in reliability of wireless communications is the problem of \emph{jamming}. This refers to the existence of a malicious user, the jammer, which tries to compromise or destroy the performance of the wireless link. We can view the problem of \emph{jamming} as a non-cooperative game \cite{OR94}. Several studies in the past studied this problem as a two-player game between the transmitter and the jammer. In particular, in \cite{AAG07} Eitan Altman et al. considered a non zero-sum non-cooperative game taking into account the transmission cost. More specifically, in this work the transmitter wants to maximize her own rate and the jammer wants to minimize the rate of the transmitter taking into account their transmission costs. The result of that work is the proof of the existence and the characterization of a unique Nash equilibrium and also an algorithm to compute it. In \cite{G09}, the author studied the problem as a zero-sum game using as utility of the transmitter the \emph{linearized Shannon capacity}. In \cite{AAG09}, the authors studied the problem in the case of multiple jammers. In \cite{JBW05} the authors studied the \emph{jamming} problem of Gaussian MIMO channels as a zero-sum game.

Inspired by the work of \cite{AAG07}, we study the problem of \emph{jamming} as a zero-sum game, with one transmitter and one jammer over multiple independent \emph{Gaussian channels} with perfect channel state information (CSI). Our scenario is a special case of the model in \cite{AAG07} when the transmission costs are zero and their results can be applied in our scenario as a subcase of their general model. Nevertheless, we believe that this scenario is of practical interest and in our work we analyze it in more detail with different and simpler tools giving simple and intuitively satisfying insights into the problem. In particular, we show that in the Nash equilibrium of the game the best-response strategy of the transmitter is the \emph{waterfilling} taking into account the sum of the jamming plus noise power in each channel, and the best-response strategy of the jammer is the \emph{waterfilling} taking into account the noise power only.

\section{The model}
We consider the \emph{Gaussian channel}, with one pair of transmitter-receiver and one malicious user, the jammer.
There are $M>0$ independent channels that the transmitter can use to transmit her information. In particular, the transmitter has a positive budget of transmission power $T$ and wants to distribute it on the channels in a way that maximizes her aggregate rate. Let the non-negative ${T_{k}}$ be the portion of the power $T$ that is used in the channel $k$, with $\sum\limits_{k=1}^{M}{T_{k}} = T$. In the absence of a jammer her optimum strategy is the well known \emph{waterfilling strategy} with respect to the channel's noise. On the other hand, the jammer has a positive budget of transmission power $J$ and wants to distribute it on the channels in a way that minimizes the aggregate rate of the transmitter. Let the non-negative $J_{k}$ be the portion of the power $J$ that is used in the channel $k$, with $\sum\limits_{k=1}^{M}J_{k} = J$. Let $\alpha_{T}>0$ be the channel attenuation for the transmitter-receiver pair equal for all channels and let $\alpha_{J}>0$ be the channel attenuation for the jammer-receiver pair also equal for all channels. Also, let $N_k>0$ be the power of the \emph{additive Gaussian white noise (AGWN)} in the channel $k$. The receiver treats the signal of the jammer as noise, so by the \emph{Shannon's formula} \cite{SH48,CT06} the rate/utility of the transmitter in nats per channel use is:
\begin{equation*}
u_T = R_T = \frac12 \sum\limits_{k = 1}^{M} \ln \left(1 + \frac{\alpha_{T}{T_{k}}}{\alpha_{J}J_{k}  + N_k}\right),
\end{equation*}
which must be maximized.

On the other hand, the utility of the jammer is:
\[u_J = u_T, \]
which must be minimized.

We can consider the transmitter and the jammer as the players in a zero-sum game \cite{OR94}. The strategies of the transmitter are the constants ${T_{k}}$ that are used to distribute her power on the $M$ channels and the strategies of the jammer are the constants $J_{k}$ that are used to distribute her power on the $M$ channels. Since, $u_T$ is concave in $T_k$ and convex in $J_k$, we can apply the Sion's minimax Theorem to conclude that it has a saddle point.

\section{The strategy of the transmitter: waterfilling}
\label{waterfilling}
We will analyse the best-response strategy of the transmitter in the zero-sum game when the strategy of the jammer is fixed, in other words the parameters $J_{k}$ are fixed. We have the following optimization problem: 
\begin{align*}
  \max_{T_k} &  \frac12 \sum\limits_{k = 1}^{M} \ln \left(1 + \frac{\alpha_T{T_{k}}}{\alpha_JJ_{k} + N_k}\right)\\ 
 & = \max_{T_k} \frac12 \sum\limits_{k = 1}^{M} \ln \left(\alpha_JJ_{k} + N_k + {\alpha_T{T_{k}}}\right)\\ & - \frac12 \sum\limits_{k = 1}^{M}\ln \left({\alpha_JJ_{k} + N_k}\right)\\
  \text{s.t.\ } 
    & \sum\limits_{k=1}^{M}{T_{k}} = {T},\\
    & {T_{k}} \geq 0, \hspace{20mm} 1\leq k\leq M.
\end{align*}
We can see that only $\frac12 \sum\limits_{k = 1}^{M} \ln \left(\alpha_JJ_{k} + N_k + {\alpha_T{T_{k}}}\right)$ depends on $T_k$ and therefore the solution of this convex optimization problem is the well known \emph{waterfilling theorem} of the Information Theory (see page 245 of \cite{BV04}) which states that $\alpha_T{T_{k}} = (v - \alpha_JJ_{k}-N_k)^{+}$, where $v$ is calculated by the expression $\sum\limits_{k=1}^{M}(v - \alpha_JJ_{k}-N_k)^{+} = \alpha_TT$. It is easy to see that $v>0$.
If for some channel(s) $j$, $\alpha_J {J_j}  + N_j \geq v$ then $T_j = 0$ and the transmitter will apply the waterfilling strategy to the rest of the channels. This situation will arise only when there is excessive noise in some channels, that is if $N_j \geq v$, since the jammer will avoid wasting power in a channel which will not be used by the transmitter. 

The waterfilling strategy of the transmitter maximizes her rate for any strategy of the jammer including the best one which minimizes this maximum. 
\section{The strategy of the jammer}
The best-response strategy of the jammer in the zero-sum game if the strategy of the transmitter is fixed, that is if  the powers $T_k$ have specific values, is determined by the following  optimization problem:
Minimize $u_T$, that is 
\begin{align*}
\label{optimization of the jammer}
\min_{J_k} & \frac12\sum\limits_{k = 1}^{M} \ln \left(1 + \frac{\alpha_T{T_{k}}}{\alpha_JJ_{k} + N_k}\right)\\
& = \min_{J_k} \frac12\sum\limits_{k = 1}^{M}\ln \left(\alpha_T {T_{k}} + \alpha_JJ_{k} + N_k\right)\\
&- \frac12\sum\limits_{k = 1}^{M}\ln\left(\alpha_JJ_{k} + N_k\right) \\
\text{s.t.\ } \\
&     \sum\limits_{k=1}^{M}J_{k} = J,\\
&     J_{k} \geq 0, \hspace{30mm} 1\leq k\leq M.\\
\end{align*}
For the analysis, we use the KKT conditions. We use a multiplier $u$ for the equation $\sum\limits_{k=1}^{M}J_{k} = J$ and a multiplier $\lambda_k$ for any condition $J_{k} \geq 0$. Thus, by the KKT conditions we have, for any $k$, the condition
\begin{equation}
\label{KKT conditions}
\frac{\alpha_J}{2(\alpha_T T_k+\alpha_J J_k+N_k )}- \frac{\alpha_J}{2(\alpha_J J_k+N_k)} + u=\lambda_k,
\end{equation} 
the complementarity slackness condition 
\begin{equation}
\label{CSC}
\lambda_k J_k = 0,
\end{equation}
and 
\[\lambda_k \geq 0.\]
From the KKT conditions when $J_k$ is positive, so $\lambda_k = 0 $, must satisfy the condition
$1/(\alpha_J J_k+N_k )-1/(\alpha_T T_k+\alpha_J J_k+N_k )=\frac{2u}{\alpha_J}$, where $\frac{2u}{\alpha_J}$ is a positive constant, since it is easy to see that $u>0$.
Solving for $J_k$  we find 
\[J_k = \frac{1}{2\alpha_J}\Big[-\alpha_TT_k-2N_k+\sqrt{(\alpha_TT_k)^2+2\frac{\alpha_J\alpha_TT_k}{u}}\Big]^+\]
and $u$ is calculated by inserting the $J_k$s into equation $\sum\limits_{k=1}^{M}J_k = J$.
We can see that $J_k$ increases as $T_k$ increases. Thus this strategy of the jammer will convert an “attractive” (= less noisy)  channel in which the transmitter applies  larger power, into a less attractive (=more noisy) channel forcing the transmitter to apply less power (assuming that she follows a waterfilling strategy) and vice versa. This best-response strategy of the jammer is also a mechanism which can be used to force the behaviour of the transmitter in a manner that leads to the Nash equilibrium of the game which is derived in the next section. 

\section{The Nash equilibrium strategies}
In this section, we will extend the previous results and observations to determine the strategies at the Nash equilibrium. In the Nash equilibrium the transmitter plays a waterfilling strategy by keeping constant the sum of her power, the power of the jammer and the noise power in each channel. The strategy of the jammer taking into account the optimal transmitter's strategy is determined by the minimax optimization problem, that is

\begin{align*}
\min_{J_k}\max_{T_k} & \frac12\sum\limits_{k = 1}^{M} \ln \left(1 + \frac{\alpha_T{T_{k}}}{\alpha_JJ_{k} + N_k}\right)\\
&= \min_{\substack{J_k\\\text{Transmitter plays}\\ \text{waterfilling}}}\frac12\sum\limits_{k = 1}^{M} \ln \left(1 + \frac{\alpha_T{T_{k}}}{\alpha_JJ_{k} + N_k}\right)\\
& = \min_{\substack{J_k\\\text{Transmitter plays}\\ \text{waterfilling}}} \frac12\sum\limits_{k = 1}^{M} \ln \left(\alpha_T {T_{k}} + \alpha_JJ_{k} + N_k\right) \\ & -\frac12\sum\limits_{k = 1}^{M}\ln\left(\alpha_JJ_{k} + N_k\right) \\
\text{s.t.\ } \\
&     \sum\limits_{k=1}^{M}J_{k} = J,\\
&     J_{k} \geq 0, \hspace{20mm} 1\leq k\leq M.\\
\end{align*}

The KKT conditions that we described in the previous section must hold at the Nash equilibrium of the game. For the analysis, we categorize the noise of a channel $k$ into three groups according to its power, $N_k\geq v$, $N_k \in (\frac{v\alpha_J}{2uv+\alpha_J},v)$ and $N_k \leq \frac{v\alpha_J}{2uv+\alpha_J}$. 
 
By the waterfilling strategy of the transmitter as we analyse in section \ref{waterfilling}, if $N_k \geq v$, then $T_k = 0$. Also, by the condition (\ref{KKT conditions}) we can see that if $T_k = 0$, then $\lambda_k >0$, so by (\ref{CSC}) we conclude that $J_k = 0$. Thus in the channels with excessive noise both the transmitter and the jammer will avoid wasting any power.

For $N_k \in (\frac{v\alpha_J}{2uv+\alpha_J},v)$, there are two cases for the jamming power: $\alpha_JJ_k \geq  v - N_k$
and $\alpha_JJ_k <  v - N_k$. The first case can not happen since then $T_k = 0 \Rightarrow  \lambda_k >0 \Rightarrow J_k =0$, but this is a contradiction. In the second case we have $\alpha_TT_k + \alpha_JJ_k+N_k=v$ and $\lambda_k >0$, since $\lambda_k = 
\frac{\alpha_J}{2v}-\frac{\alpha_J}{2(\alpha_JJ_k + N_k)} +u \geq  \frac{\alpha_J}{2v}-\frac{\alpha_J}{2N_k} +u>\frac{\alpha_J}{2v} -\frac{(2uv+\alpha_J)}{2v} +u = u-u= 0 \Rightarrow J_k = 0$. Thus the channels with $N_k \in (\frac{v\alpha_J}{2uv+\alpha_J},v)$ will be used by the transmitter but not by the jammer since the noise is large enough to make jamming inefficient and a waste of jamming power.

For $N_k \leq \frac{v\alpha_J}{2uv+\alpha_J}$, the only possibility for the jamming power is $\alpha_JJ_k <  v - N_k$ (for the same reason as above). In this case we have $\alpha_TT_k +\alpha_JJ_k+N_k = v$ and $\lambda_k = 0$, since the case $\lambda_k>0 \Rightarrow J_k = 0$ leads to the contradiction $N_k > \frac{v\alpha_J}{2uv+\alpha_J}$. Setting $\alpha_TT_k +\alpha_JJ_k+N_k = v$ and $\lambda_k = 0$ in (\ref{KKT conditions}) and solving for $J_k$ we find that $\alpha_JJ_k = \frac{v\alpha_J}{\alpha_J + 2uv} - N_k$. The constant $u$ can be found by solving the equation $\sum\limits_{k=1}^{M}(\frac{v\alpha_J}{\alpha_J + 2uv} - N_k)^+ = \alpha_JJ$.
The preceding analysis proves that the strategy of the jammer in the Nash equilibrium is also the waterfilling with respect to the noise power of the channels.

In particular, if the powers of the transmitter and the jammer are much larger than the noise power then the waterfilling strategy of the jammer makes the combined jamming plus noise power equal in all channels whereas the waterfilling strategy of the transmitter results in the uniform distribution of her power in all channels.

\section{Conclusions}
In this paper we study the problem of \emph{jamming} in multiple independent \emph{Gaussian channels}. We derive the strategies of the transmitter and the jammer on the unique Nash equilibrium where the transmitter maximizes the minimum of her rate and the jammer minimizes the maximum of the transmitter rate. In particular, given that the waterfilling strategy of the transmitter is known as the best strategy under interference, our main contribution in this paper is to show that the jammer's optimum strategy in the Nash equilibrium is the waterfilling to the noise of the channels as well.

\bibliographystyle{plain}
\bibliography{agt}

\end{document}